# Unobtrusive Reflectance Photoplethysmography for Detecting and Severity Grading of Sleep Apnea via Oxygen Desaturation Index

Karen Adam*, Clémentine Aguet, Patrick Theurillat, Florent Baty, Maximilian Boesch,

Damien Ferrario, Mathieu Lemay, Martin Brutsche, Fabian Braun

**Abstract:** Sleep apnea is a common chronic sleep-related disorder which is known to be a comorbidity for cerebro- and cardio-vascular disease. Diagnosis of sleep apnea usually requires an overnight polysomnography at the sleep laboratory. In this paper, we used a wearable device which measures reflectance photoplethysmography (PPG) at the wrist and upper arm to estimate continuous $SpO_2$ levels during sleep and subsequently derive an oxygen desaturation index (ODI) for each patient. On a cohort of 170 patients undergoing sleep apnea screening, we evaluated whether this ODI value could represent a surrogate marker for the apnea-hypopnea index (AHI) for the diagnosis and severity assessment of sleep apnea. As the ODI was simultaneously obtained at the fingertip, upper arm and wrist, we compared ODI diagnostic performance depending on the measurement location. We then further evaluated the accuracy of ODI as a direct predictor for moderate and severe sleep apnea as defined by established AHI thresholds. We found that ODI values obtained at the upper arm were good predictors for moderate or severe sleep apnea, with 86% accuracy, 96% sensitivity and 70% specificity, whereas ODI values obtained at the wrist were less reliable as a diagnostic tool.

**Keywords:** Sleep apnea, Photoplethysmography, Pulse oximetry.

*****Corresponding author: Karen Adam:** Centre Suisse d'Electronique et de Microtechnique (CSEM), Neuchâtel, Switzerland, karen.adam@csem.ch.
**Clémentine Aguet, Patrick Theurillat, Damien Ferrario, Mathieu Lemay, Fabian Braun:** Centre Suisse d'Electronique et de Microtechnique (CSEM), Neuchâtel, Switzerland.
**Florent Baty, Maximilian Boesch, Martin Brutsche:** HOCH Health Ostschweiz, Kantonsspital St.Gallen, Lung Center, St.Gallen, Switzerland**.**

## 1 Introduction

Sleep apnea (SA) is a chronic sleep-related disorder which consists of repetitive pauses or restrictions in airflow when breathing. SA is a known risk factor for cerebro- and cardio-vascular disease and affects an estimated 22.6% of the population [1]. SA is generally diagnosed following a full-night polysomnography (PSG) at a hospital or dedicated sleep center. The severity of SA can be assessed by examining an apnea-hypopnea index (AHI) which measures the average number of apneic events per hour of sleep [2]. Each apneic event may result in a drop in blood oxygen levels which itself may produce corresponding drops in measured peripheral blood oxygen saturation ($SpO_2$). Therefore, the oxygen desaturation index (ODI) obtained from observing drops in $SpO_2$ values in a patient, can be used as surrogate marker to indicate SA severity.

In this paper, we evaluate the feasibility of using wearable reflectance photoplethysmography (PPG) devices placed at the upper arm or wrist to detect moderate or severe SA via the ODI. Using such devices would allow a SA severity grading over multiple nights, whereas SA detection based on a single night's PSG can be highly influenced by inter-night variability in SA severity, leading to over- and undertreatment if only one night is considered for diagnosis [1].

# 2 Methods

## 2.1 Data Collection

Subjects suspected of having SA underwent overnight PSG as part of the routine diagnostic workup. During this occasion, reflectance PPG signals were acquired at the wrist and upper arm using a CSEM-proprietary wearable device (Figure 1). The data in question are a subset of a wider data collection campaign taking place at the HOCH Health Ostschweiz which encompasses 700 subjects, and which was approved by the local ethics committee (Ethikkommission Ostschweiz, BASEC-ID: 2024-00637).

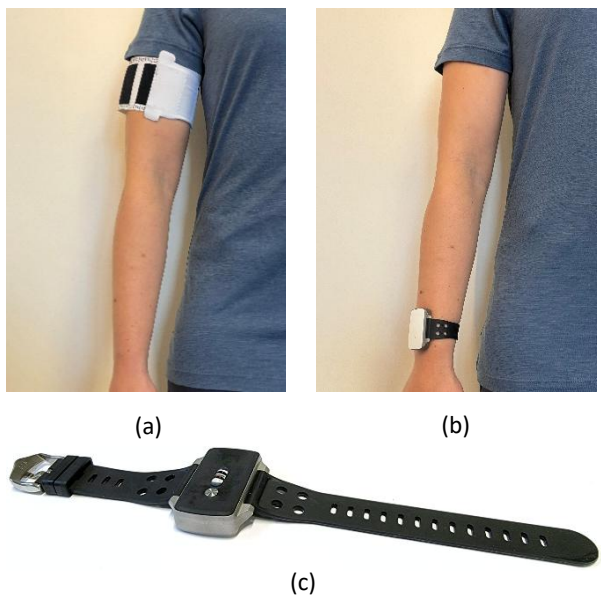

**Figure 1:** CSEM wearable worn at (a) upper arm and (b) wrist and (c) back view (skin-facing side) of wearable.

At the time of publication, data from 170 of these subjects were available (see Table 1 for population information).

For each of these subjects, we had access to a reference $SpO_2$ value from the fingertip PPG (Xpod 3012LP, Nonin Inc., Plymouth, USA), as well as an AHI which is scored by the PSG.

We furthermore had access to red, infrared and green reflectance PPG signals acquired using the wearable PPG devices placed at the upper arm and wrist. These signals allow us to obtain an estimate of $SpO_2$ at each of the two measurement locations. We have studied, in a previous publication, the effect of the measurement location on measurement accuracy and signal acceptance rate for a subset of the data considered here [3].

**Table 1:** Cohort statistics (N=170).

| Characteristics | Median (or count) | IQR (or percentage) |
|---|---|---|
| Age (yrs) | 50 | 38-60 |
| Gender, male (count) | 125 | 73.1 |
| Body Mass Index (kg/m2) | 29 | 26.3-33.1 |
| Apnea-Hypopnea Index (events/hour) | 27 | 11-49.5 |
| Normal SA, AHI ≤ 5 (count) | 21 | 12.4 |
| Mild SA, 5 < AHI ≤ 15 (count) | 36 | 21.2 |
| Moderate SA, 15 < AHI ≤ 30 (count) | 35 | 20.6 |
| Severe SA, AHI > 30 (count) | 78 | 45.9 |

## 2.2 Processing Setup

The PPG signals recorded at the upper arm and wrist were acquired at 50 Hz for both the red and infrared channels and at 100 Hz for the green channel. The data were processed using a CSEM proprietary $SpO_2$ estimation algorithm which returns $SpO_2$ values with an averaging window length of 3 seconds and a refresh rate of 1 second. For each $SpO_2$ value computed over a window, we also obtained a corresponding quality index value which indicates trust in the $SpO_2$ estimates.

Sections of the signal where the quality index was lower than an empirically derived threshold were set to zero and treated as artefacts for the remainder of the processing pipeline. Following this processing, we had access to three $SpO_2$ traces from different measurement locations: 1) a reference measured using transmittance PPG at the fingertip; 2) an estimate measured at the upper arm using reflectance PPG; and 3) an estimate measured at the wrist using reflectance PPG.

For each of these $SpO_2$ traces, we obtained an ODI using the ABOSA software [4]. We set a threshold of 3% for drops in $SpO_2$ levels to detect apneic events, as recommended by the American Association of Sleep Medicine (AASM) [2]. Furthermore, we classified apneic events only when they lasted longer than 5 seconds.

We assessed the agreement, i.e. bias and standard deviation, between the resulting ODIs by comparing the ODI derived from fingertip reference transmittance $SpO_2$ vs. ODIs derived from reflectance $SpO_2$ measured using the wearables at the wrist and upper arm, respectively.

For each of the three measurement locations, we used the ODI as a regressor for a reference AHI value computed by the PSG methodology. Moreover, we searched for a SA classifier based on the ODI value. To this end, we first considered the four common classes of SA based on standard AHI thresholds: normal (AHI < 5/h), mild (5 ≤ AHI < 15/h), moderate (15 ≤ AHI < 30/h), and severe (AHI ≥ 30/h). Based on these categories, we evaluated the ability in detecting three different severities of SA (AHI > 5; AHI > 15; AHI > 30) by applying a simple ODI threshold. For each measurement location and severity class, ODI thresholds were chosen as the value which maximizes the sum of the squared sensitivity and squared specificity. Using these thresholds, binary classification performance was evaluated in terms of sensitivity, specificity, and accuracy.

## 3 Results and Discussion

Table 2 lists the accuracy of wearable-derived $SpO_2$ estimation with respect to the measurement location when compared to fingertip reference values. In brief, the $SpO_2$ estimates at the upper arm yielded a lower error and higher acceptance rate compared to the wrist. These findings are in line with previously reported results on a smaller subset of the study population [3].

**Table 2:** $SpO_2$ estimation accuracy vs. measurement location.

| Measurement Location | Error Bias (%) | Error $A_{RMS}$ (%) | Acceptance Rate (%) |
|---|---|---|---|
| Upper Arm | 0.4 | 2.0 | 96.0 |
| Wrist | -0.9 | 3.5 | 68.2 |

Comparison of the ODI derived from fingertip reference $SpO_2$ vs. the other two measurement locations yielded a bias [95% confidence interval] of 3.92 [-5.12, 17.44] events/hour for the upper arm and 7.35 [-22.1, 31.07] events/hour for the wrist, respectively.

The top row of Figure 2 depicts the relationship between estimated ODI and reference AHI provided by the PSG for each measurement location. The bottom row of Figure 2 shows the receiver operating characteristic (ROC) curves obtained when using ODI to detect moderate or severe SA (i.e. classifying subjects with AHI>15 events/hour). Table 3 lists the performance for each measurement location for classifying SA severity. We noticed a decrease in overall accuracy when SA severity increases for all measurement locations. We conjecture this to be due to the high prevalence of moderate or severe SA in our dataset. For all three SA severity classes, the fingertip-based classification yielded the best results, closely followed by the values derived at the upper arm, indicating that the upper arm might be a suitable location for ODI estimation using wearables. In contrast, when assessing $SpO_2$ at the wrist, the classification performance was lower, which may also be explained by the higher error of $SpO_2$ and ODI estimation at this measurement location when compared to the reference fingertip. Thus, future work should investigate the reasons for lower performance for $SpO_2$ and ODI estimation at the wrist, a common location for assessing vital signs using smartwatches or bracelets.

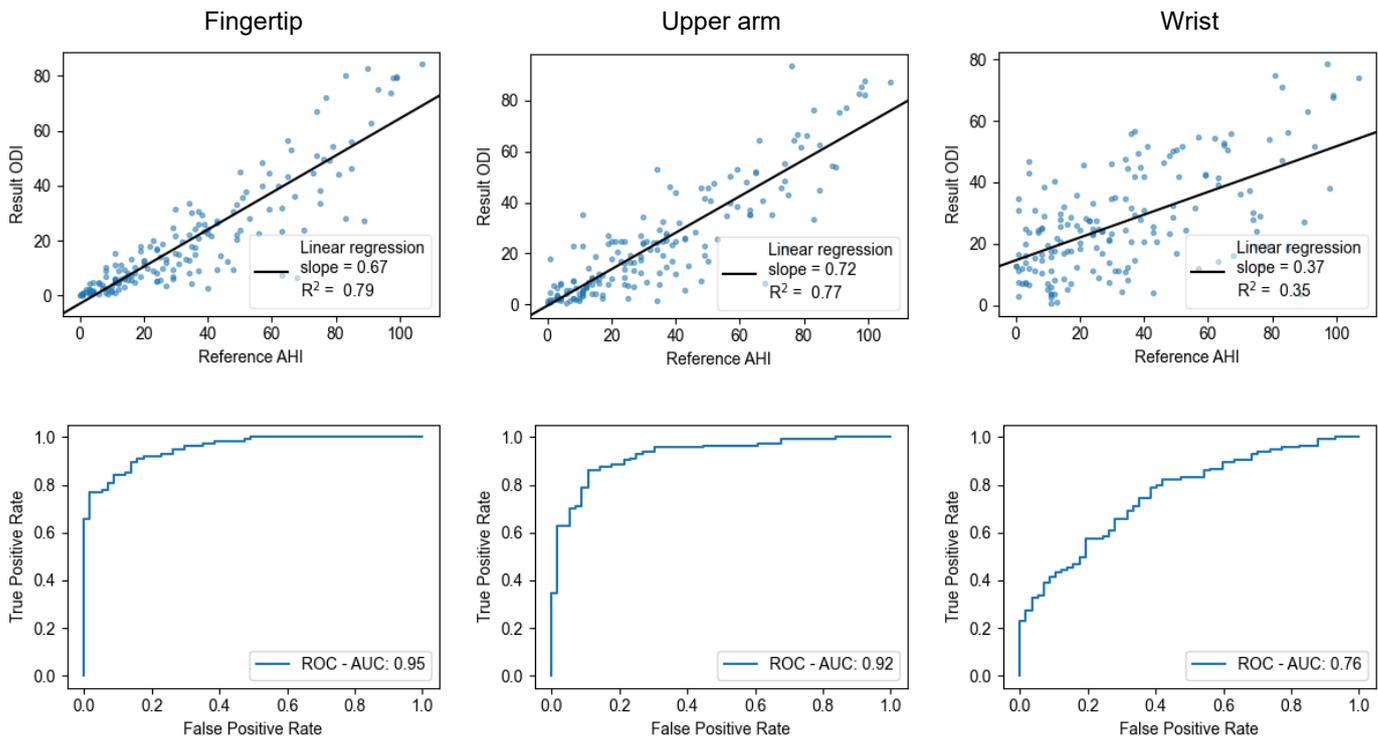

**Figure 2:** (Top) Plots of oxygen desaturation index (ODI) estimated from transmittance PPG at the fingertip (Left), reflectance PPG at upper arm (Middle) and reflectance PPG at wrist (Right) compared to reference apnea hypopnea index (AHI) computed by PSG, along with linear regression curve. (Bottom) Receiver operating characteristics (ROC) curves obtained when using ODI for classifying moderate or severe apnea (AHI>15 events/hour).

**Table 3:** Performance to classify three different sleep apnea severities using the oxygen desaturation index (ODI) derived from three measurement locations: fingertip, upper arm, and wrist.

| Sleep Apnea Severity | Measurement Location | AUC (-) | ODI Threshold (events/hour) | Sensitivity (%) | Specificity (%) | Accuracy (%) |
|---|---|---|---|---|---|---|
| Mild, moderate or severe (AHI > 5) | Fingertip | 0.97 | 2.3 | 96 | 86 | 94 |
| | Upper Arm | 0.91 | 3.7 | 93 | 70 | 89 |
| | Wrist | 0.70 | 14.1 | 80 | 57 | 76 |
| Moderate or severe (AHI > 15) | Fingertip | 0.95 | 5.0 | 96 | 70 | 87 |
| | Upper Arm | 0.92 | 7.1 | 96 | 70 | 86 |
| | Wrist | 0.76 | 17.0 | 82 | 58 | 74 |
| Severe (AHI > 30) | Fingertip | 0.94 | 7.5 | 96 | 64 | 78 |
| | Upper Arm | 0.91 | 11.9 | 92 | 69 | 79 |
| | Wrist | 0.77 | 19.2 | 80 | 52 | 64 |

The generalizability of our work is limited by the single use of ODI for classifying SA. In the future, this detection will be enhanced by using a variety of other PPG-derived features (including heart rate variability or respiratory rate) which are expected to improve the performance, thus allowing an accurate and unobtrusive detection of SA using wearable devices.

## 4 Conclusion

We here evaluated the use of ODI, retrieved from different measurement as a diagnostic tool for mild, moderate and severe SA, on a cohort of 170 subjects who underwent an overnight PSG at a tertiary care center in Switzerland. We showed that ODI estimates obtained at the upper arm can predict moderate to severe SA with an accuracy of 86%, a sensitivity of 96% and a specificity of 70%. Such a classifier fares better than a classifier for moderate to severe sleep apnea based on wrist ODI. In the future, we will incorporate other PPG-derived features such as heart rate variability or respiratory rate, enabling accurate and unobstructive detection of SA using wearables.

**Author Statement**
Research funding: This work was supported by the Swiss National Science Foundation (SNF) project ApneaSense (205321_219441).